\documentclass{article}
\usepackage{spconf,amsmath,graphicx}
\usepackage{caption}
\usepackage{subcaption}
\usepackage{makecell}
\usepackage{amssymb,amsfonts}
\usepackage{lipsum}
\usepackage{graphicx}
\usepackage{colortbl}
\usepackage{enumerate}
\usepackage{enumitem}
\usepackage{multirow}
\usepackage{textcomp}
\usepackage{textpos}

\title{Quality and Complexity Assessment of Learning-Based Image Compression Solutions}
\name{João Dick$^1$, Brunno Abreu$^1$, Mateus Grellert$^2$, Sergio Bampi$^1$\thanks{The authors thank CNPq, CAPES Finance code 001, and FAPERGS Brazilian Agencies for R\&D support for scholarships and  financial support.}}
\address{$^1$ Informatics Institute (PGMICRO), Federal University of Rio Grande do Sul, Porto Alegre, Brazil \\
$^2$ Graduate Program in Computer Science, Federal University of Santa Catarina, Florian\'opolis, Brazil}
\begin{document}
\maketitle
\begin{abstract}
This work presents an analysis of state-of-the-art learning-based image compression techniques. We compare 8 models available in the Tensorflow Compression package in terms of visual quality metrics and processing time, using the KODAK data set. The results are compared with the Better Portable Graphics (BPG) and the JPEG2000 codecs. Results show that  JPEG2000 has the lowest execution times compared with the fastest learning-based model, with a speedup of 1.46$\times$ in compression and 30$\times$  in decompression. However, the learning-based models achieved improvements over JPEG2000 in terms of quality, specially for lower bitrates. Our findings also show that BPG is more efficient in terms of PSNR, but the learning models are better for other quality metrics, and sometimes even faster. The results indicate that learning-based techniques are promising solutions towards a future mainstream compression method. 
\end{abstract}
\begin{textblock}{13.25}[0,0](0,6.35)
\footnotesize© 2021 IEEE. Personal use of this material is permitted. Permission from IEEE must be obtained for all other uses, in any current or future media, including reprinting/republishing this material for advertising or promotional purposes, creating new collective works, for resale or redistribution to servers or lists, or reuse of any copyrighted component of this work in other works.
\end{textblock}


\section{Introduction}
\label{sec:intro}

The demand for multimedia services has experienced a huge increase in the latest years, making image and video content account for most of the web's data traffic \cite{cisco2018cisco}. To address this issue, compression solutions are of utmost importance. When compressing an image, the main goal is to reduce the image file size as much as possible with minimal loss of  quality, which has to be quantified by a particular visual quality metric. Numerous algorithms have been proposed over the years combining transforms, operations, and hand-tuned parameters to accomplish suitable rate-distortion trade-offs. These classical models include JPEG2000 \cite{jpeg2000} and Better Portable Graphics (BPG) \cite{BPG}, which is based on the intra-frame prediction of the High Efficiency Video Coding standard \cite{hevc}.

Learning-based image compression is a new paradigm that has become possible due to advances in hardware computational performance, more sophisticated algorithms, and an easier online access to a massive amount of training data. This modern approach is based in the use of Convolutional Neural Networks (CNNs) and manipulates data compression in a different way: the majority of compression parameters is no longer hand-tuned, but learned in a training process that aims at minimizing a rate-distortion curve. Enormous improvements in this field have been observed in the latest years through several contributions \cite{balle2018efficient, balle2018variational, minnen2018joint, mentzer2020highfidelity}. Quality analyses of learning-based image compression approaches are already available in the literature \cite{cheng2019perceptual, valenzise:hal-01819588}. However, they lack an execution time analysis of the solutions, which is essential to understand the applicability of these novel methods. Additionally, the data sets employed are not common in the literature. 
Lastly, the models of these references are outdated, so a more recent state-of-the-art analysis is required. 

The goal of this work is to present a detailed overview on the current learning-based image compression models in the literature, followed by a systematic comparison amongst them. We herein compare the models in terms of several metrics, using the KODAK \cite{kodak} data set.
The contributions of this work are two-fold: i) detailed comparisons between 8 different models of recent learning-based image compression solutions, in terms of three quality metrics -- Peak Signal-to-Noise Ratio (PSNR), Multiscale Structural Similarity (MS-SSIM) and Learned Perceptual Image Patch Similarity (LPIPS) \cite{lpips} is presented; and ii)  comparisons in terms of execution time of those models is also provided. Additionally, execution time comparisons between the models running on CPU and GPU in a low-cost hardware are also presented.



\section{Learning-Based Image Compression}
\label{sec:models}
Regardless of the architecture, compression models follow a common scheme. In order to compress an image $x$, an encoder $E$ is designed to transform it into a latent representation $y=E(x)$. The latent data is then quantized to $\hat{y}=Q(y)$, resulting in a discrete-valued data vector. The discrete representation can be compressed by an entropy coder along with a probability or entropy model $P$. After this step, the image achieves its most compressed form and can be stored or transmitted as a bit stream. To reconstruct the image, the bit stream is sent to a decoder $\ G$, yielding  $\ x'=G(\hat{y})$, with distortion measured by $\ d(x,x')$ according to a given quality metric.

In learning-based schemes, $P$, $G$ and $E$ are trained as CNNs minimizing a loss function $L=R(\hat{y})\lambda + d(x,x')$. This function is a combination of quality ($d$) and compression ($R$), and it has a great influence in the training process, since all the network parameters are optimized according to a trade-off between these terms.

\begin{figure}
    \centering
    \includegraphics[width=1\linewidth]{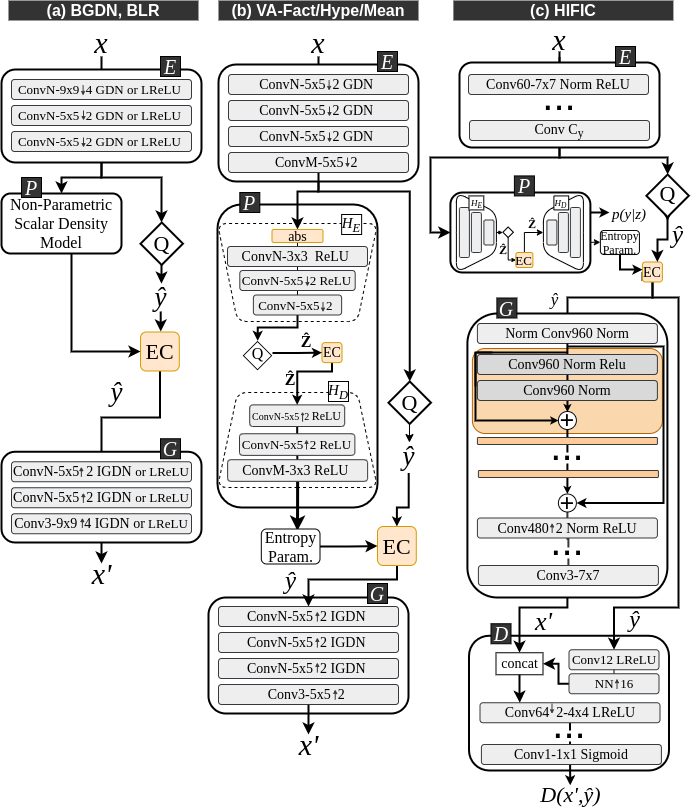}
    \caption{CNN Architectures used to train the eight models evaluated in this work. EC stands for Entropy Coding.}
    \label{fig:models}
    \vspace{-0.5cm}
\end{figure}

Our analysis covers eight pre-trained compression models available in the Tensorflow \cite{abadi2016tensorflow} Compression library. All models are reproductions of the ones used in \cite{balle2018efficient, balle2018variational, minnen2018joint, mentzer2020highfidelity}, and only differ in using integer arithmetic on operations related with conditional priors. This technique is described in \cite{balle2019integer} and provides consistent performance across different hardware platforms. Hence, the results presented in our work may have a minor difference from those published by the authors. Fig. \ref{fig:models} shows the CNN architectures used to train the eight models evaluated in this work, and the following paragraphs give a brief description of each. 

\textbf{BLR and BGDN: }the work of \cite{balle2018efficient} presents different architectures of nonlinear transform coders. The models are built with factorized priors and trained for the mean squared error quality metric. A comparison is made when switching the activation function from leaky ReLU (Rectified Linear Unit) to Generalised Divisive Normalization (GDN). GDN performs local normalization, hence decreasing statistical dependencies. The work also analyzes a variation in the number of filters, using $N$ =128 and  $N$ = 192 filters. This parameter was set with 4 different $\lambda$ values generating 4 learning-based models named as BLR-128, BLR-192, BGDN-128 and BGDN-192. The model's architecture is described Fig. \ref{fig:models}(a).

\textbf{VA-Fact/Hype/Mean:} the work presented in \cite{balle2018variational} proposes a Variational Autoencoder (VA) architecture that explores the use of side information sent from the encoder $E$ to the decoder $G$, enabling improvements in the entropy model. The mismatch between the rate and distortion distributions is minimized with a hyperprior architecture. This method adds another encoder-decoder pair $H_{E}-H_{D}$ used to learn hyper-latents: compressed data regarding the spatial distribution of standard deviations $\hat{\sigma}$ from $y$. $\hat{\sigma}$ is used to capture spatial dependencies in $y$ yielding to more efficient entropy coding (EC). Two different models will be addressed in our analysis: the first consisting of $E$ and $G$, named as VA-Factorized (VA-Fact), and a second one with the addition of $H_{E}$ and $H_{D}$ called VA-Hyperprior (VA-Hype). In \cite{minnen2018joint}, different combination of priors are explored. It extends the architecture from \cite{balle2018variational} by generalizing a Gaussian scale mixture (GSM) into a conditional Gaussian mixture model (GMM). We evaluated the hyperprior model with non zero-mean Gaussian conditionals, referred in our work as VA-MEAN. Fig. \ref{fig:models}(b) shows the CNN architecture used in these works.

\textbf{HiFiC:} Generative Adversarial Network (GAN) based models are presented in \cite{mentzer2020highfidelity}. Conditional GANs are designed to learn a generative model of a conditional distribution $p_{X|S}$. Each data point $x$ from $X$ is related to additional information $s$ trough an unknown joint distribution $p_{X,S}$. The authors train two networks composed by a generator $G$ conditioned in $s$ to map samples $y$ from a known distribution $p_{Y}$ to $p_{Y|S}$. A discriminator $D$ maps an input (x,s) to its probability of being a sample from $p_{X|S}$ instead of being generated by $G$. The GAN is combined with the hyperprior model used in \cite{balle2018variational}, resulting in the HiFiC network displayed in Fig. \ref{fig:models}(c).


 
\section{Methodology}
\label{sec:methodology}
\begin{figure}
    \centering
    \includegraphics[width=0.85\linewidth]{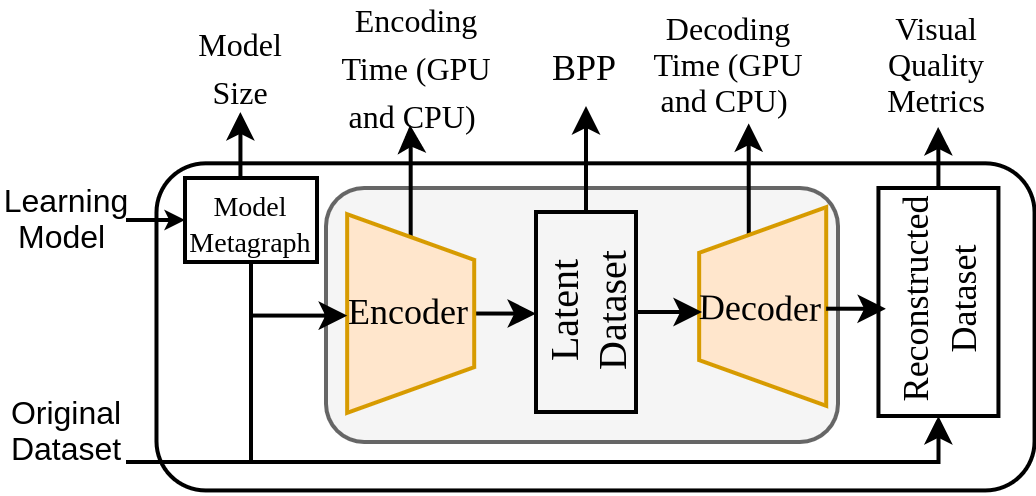}
    \caption{Execution flow used to evaluate each model.}
    \label{fig:flow}
        \vspace{-0.5cm}
\end{figure}

The execution flow used in this work is presented in Fig. \ref{fig:flow}. We evaluated the eight learning based models, comparing them with JPEG2000, using the OpenJPEG \cite{openjpeg} implementation, and BPG. For models trained under more than one distortion metric, we opted to use only the ones based on the Mean Square Error (MSE). The chosen data set and model metagraph (file containing parameters) are fed into the encoder model yielding the latent data set, model size, encoding time, and Bits Per Pixel (BPP), a commonly adopted metric for compression, as lower BPP values represent more compression. In the decoding process, the latent data set is reconstructed as images, allowing us to obtain the decoding time and the visual metrics results -- PSNR, MS-SSIM and LPIPS. We evaluate both PSNR and MS-SSIM in the luminance (Y) channel of the YCbCr color space. PSNR values were computed with the SEWAR Python Package
. The LPIPS metric is specifically used in deep visual representations \cite{lpips}, and it is evaluated using RGB due to the metric standards. 

The simulation server uses an Intel Xeon 3.6GHz CPU, with an NVidia GeForce GTX 1080 (8GB), with 32GB of RAM memory, and Cent OS 6. The Python scripts use Tensorflow-gpu 1.15 and tensorflow-compression 1.3 versions. The KODAK \cite{kodak} data set with 24 768×512 RGB images was used as benchmark, with a bit depth of 8.

One of the challenges of comparing the compression models is that each one has its own operating points in the BPP-quality curve, so simply comparing the means is not a fair approach. Thus, we computed the mean performance under a common BPP interval. After such intervals were found, we generalized each BPP-quality curve using a Piecewise Hermite Cubic Interpolator (PCHI) \cite{kreyszig2009advanced}. 
We used this approach to compute the PSNR and MS-SSIM means of all models, except for HiFiC due to its limited operation range. The results of these assessments are presented in the next section.



\section{Results and discussions}
\label{sec:results}

This section will evaluate quality and complexity using the results obtained with the methodology previously described, averaged across all 24 images of the KODAK data set.

\vspace{-0.25cm}
\subsection{Quality Analysis}
\begin{figure*}[htb]
    \begin{subfigure}[b]{0.33\textwidth}
        \includegraphics[width=\textwidth]{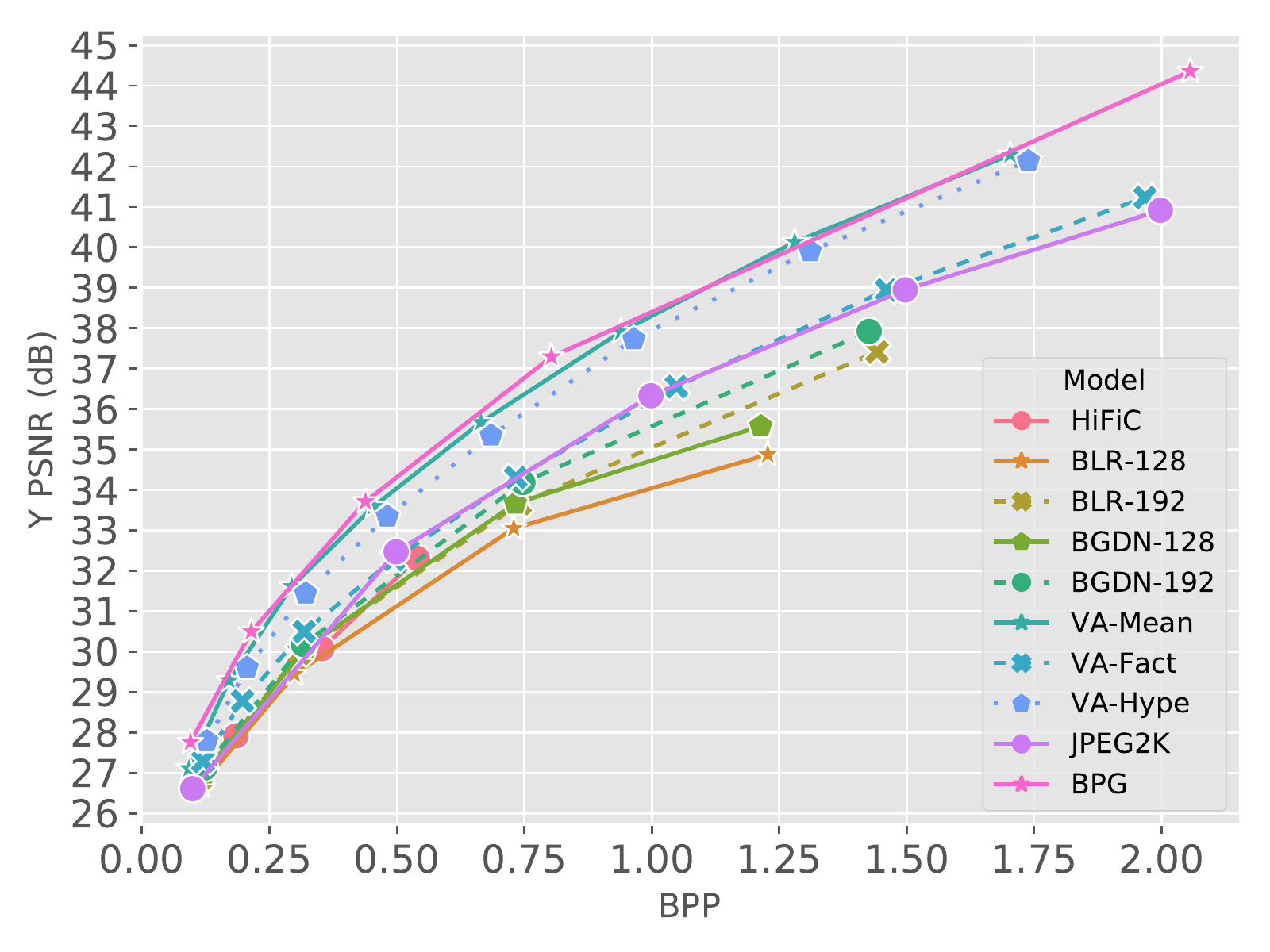}
            \vspace{-0.65cm}
        \caption{Y PSNR}
    \end{subfigure}
    \begin{subfigure}[b]{0.33\textwidth}
        \includegraphics[width=\textwidth]{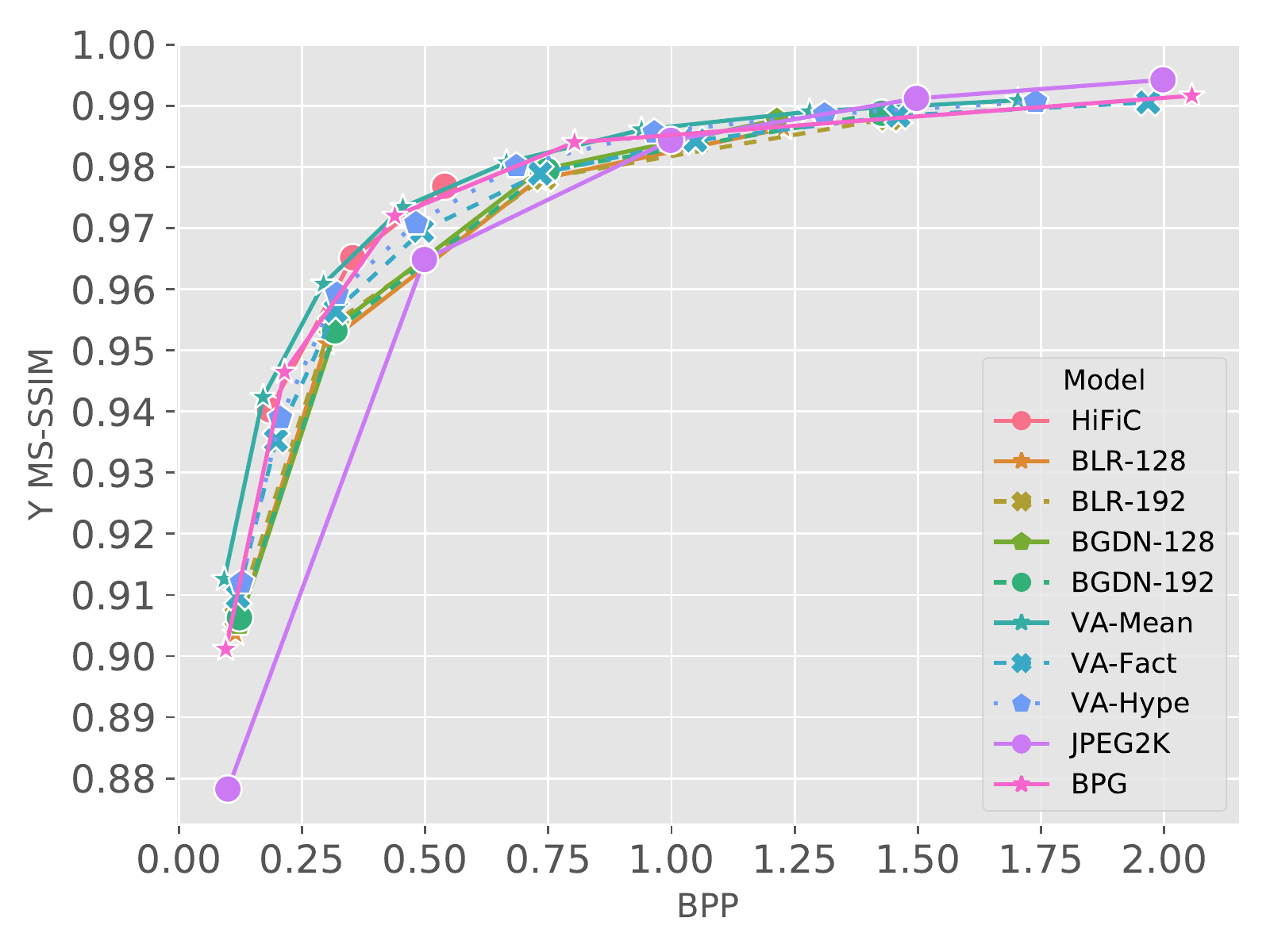}
            \vspace{-0.65cm}
        \caption{Y MS-SSIM}
    \end{subfigure}
    \begin{subfigure}[b]{0.33\textwidth}
        \includegraphics[width=\textwidth]{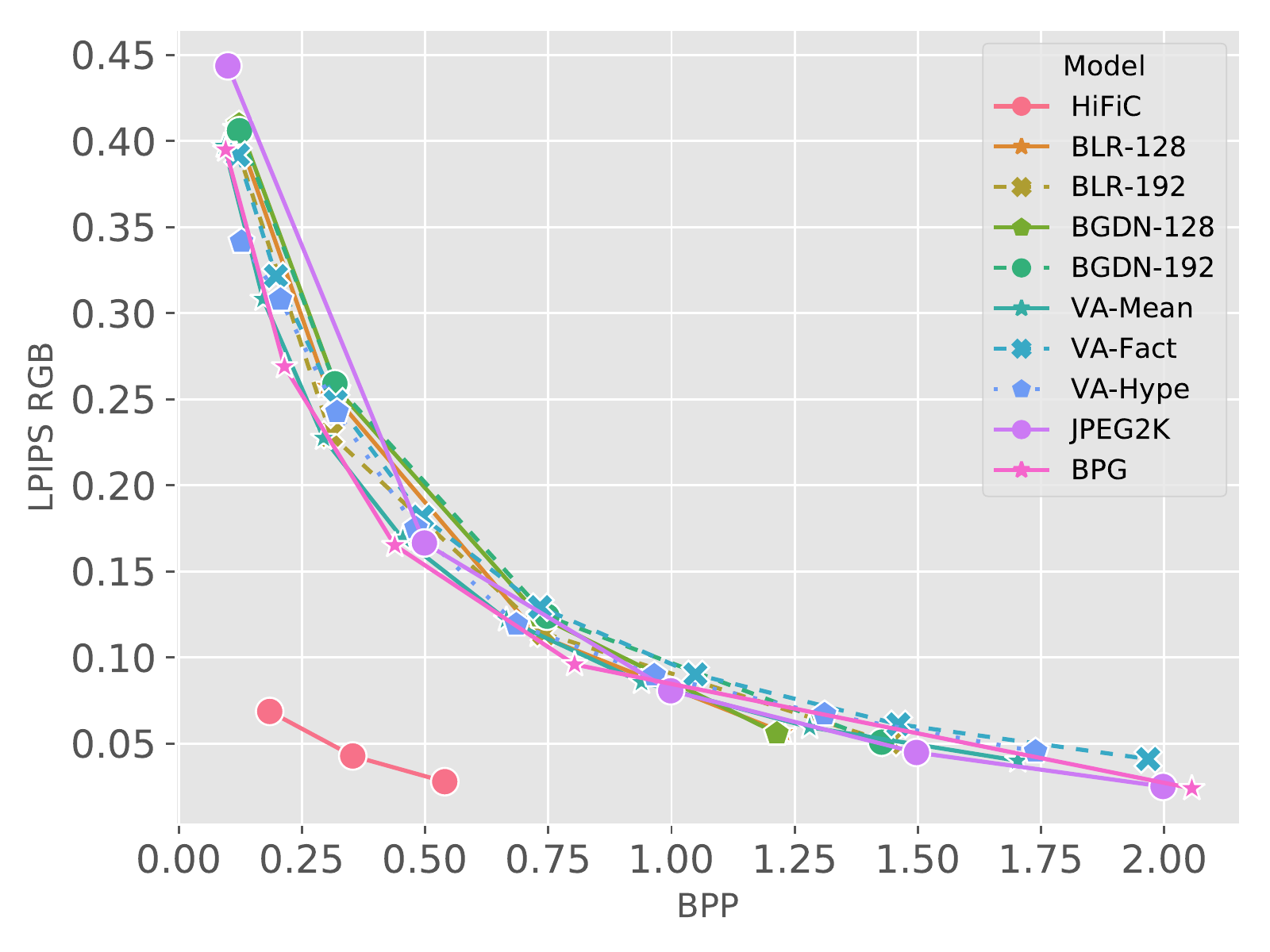}
            \vspace{-0.65cm}
        \caption{RGB LPIPS}
    \end{subfigure}
        \vspace{-0.65cm}
    \caption{Objective metrics comparison of the studied solutions, reporting (a) PSRN, (b) MS-SSIM, and (c) LPIPS for different compression levels (in bits per pixel - BPP).}
    \label{fig:psnr_ssim_lpips}
\end{figure*}

The results presented in Fig. \ref{fig:psnr_ssim_lpips} show the 10 models evaluated (JPEG2000, BPG along with 8 learning-based models) with the three quality metrics discussed in the previous section: PSNR, MS-SSIM, and LPIPS.

Most tested models achieved a broad range of operating points in the compression-quality space, as shown in Fig. \ref{fig:psnr_ssim_lpips}. This is advantageous for adaptive compression applications that provide more quality at the cost of higher bitrates. BPG and the VA model family showed the best overall results in both PSNR and MS-SSIM metrics, whereas the HiFiC family achieved better results for its target metric (LPIPS). However, the performance gap between HiFiC and the others is significantly larger in Fig. \ref{fig:psnr_ssim_lpips}(c), so if the best results among all metrics are considered, HiFiC has the advantage. The main drawback of this family is its limited operating range, with only three points, which makes it difficult to compare it with the others in terms of average performance. 
The BPG codec is superior in terms of PSNR across all the BPP range, but this advantage dwindles when the other two metrics are observed, indicating that learning-based models are likely better in terms of subjective quality. 

\begin{figure}[!t]
    \centering
    \includegraphics[width=0.95\linewidth]{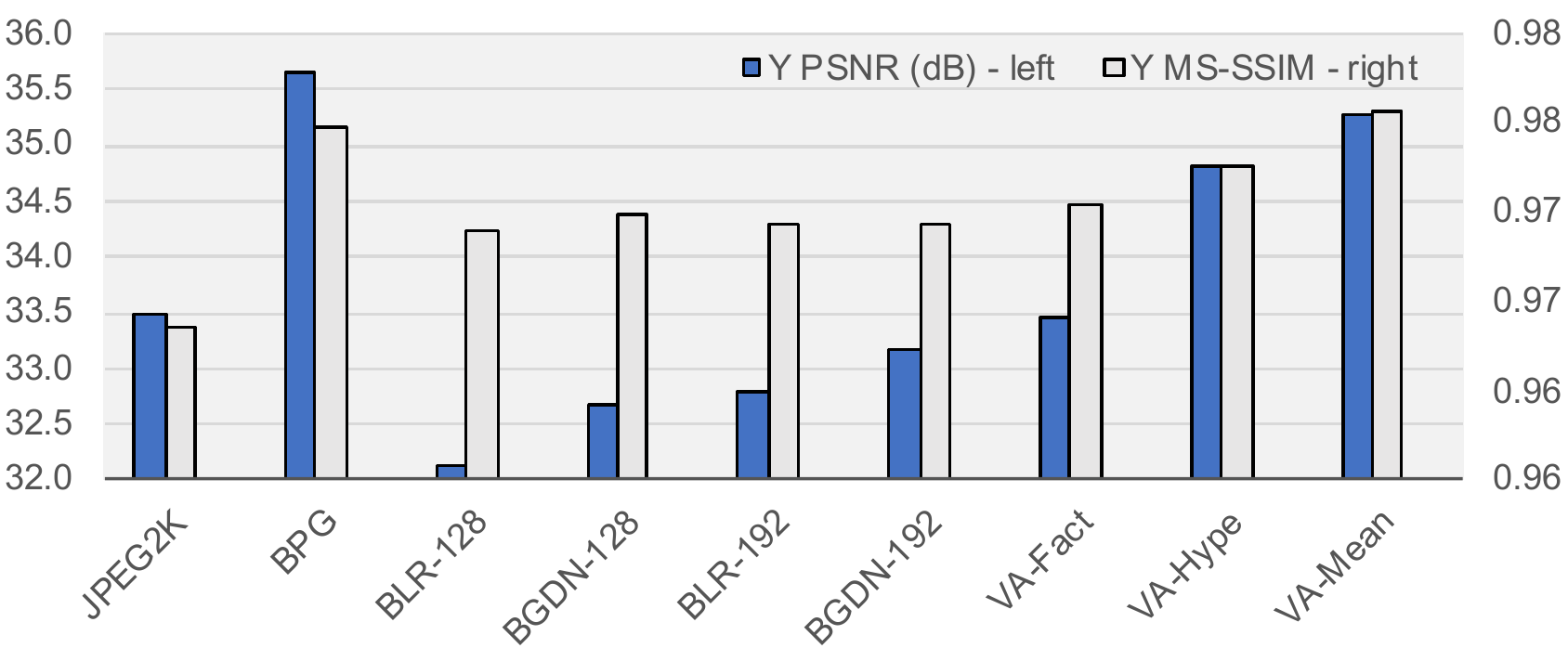}
    \caption{Mean Y PSNR and MS-SSIM values of each model, considering common BPP intervals and cubic interpolation.}
    \label{fig:overall_psnr_ssim}
        \vspace{-0.5cm}
\end{figure}

The analysis depicted in Fig. \ref{fig:overall_psnr_ssim} measured the mean PSNR and MS-SSIM values of each model family under a common BPP interval. This allows us to figure out which solution performs better overall while keeping a fair compression range among all models. To do so, 
a cubic interpolator was used to model a continuous (BPP, quality) mapping of each. Due to its limited operating range, the HiFiC model was left out of this experiment. The results show that, despite the fact that JPEG2000 presents competitive results when using PSNR (only being outperformed by the VA models), it presents the worst average results in terms of MS-SSIM. This confirms what was seen in our previous analysis in Fig. \ref{fig:psnr_ssim_lpips}(b), which showed that JPEG2000 presented significantly worse results in lower BPPs w.r.t. the remaining models, which translated to the worst average in Fig. \ref{fig:overall_psnr_ssim}. BPG outperforms all models for PSNR, and is only surpassed by VA-Mean for MS-SSIM. This indicates that the learned solutions still require improvement to justify their adoption, but we must keep in mind that only the models trained for MSE were considered in our analysis, not for MS-SSIM. This analysis also shows that the VA-Hype and VA-Mean present the most balanced results in terms of both metrics.

\vspace{-0.25cm}
\subsection{Complexity Analysis}

We also collected the compression and decompression processing times under both CPU and GPU execution platforms. The average results are presented in Tab.~\ref{tab:processing_times}, along with the average model size of each family.

\begin{table}[!b]
    \vspace{-0.25cm}
\caption{Model size and mean compression/decompression times of each model running on CPU and GPU platforms}
    \vspace{-0.2cm}
\label{tab:processing_times}
\centering
\small
\begin{tabular}{cc|cc|cc}
\hline
 & & \multicolumn{2}{|c}{Comp. time (s)} &  \multicolumn{2}{|c}{Dec. time (s)} \\\hline
Model                   & \makecell{Size\\(MB)} &\makecell{CPU} & \makecell{GPU} &\makecell{CPU} & \makecell{GPU}\\
\hline
JPEG2K     &  --   & 0.13  & --   & 0.01 & -- \\
BPG     &  --   & 0.24  & --   & 0.16 & -- \\
BLR-128    & 6.6   & 0.19 & 0.18 & 0.30 & 0.24 \\
BGDN-128.  & 6.9  & 0.20 & 0.18 & 0.31 & 0.23 \\
BLR-192.   & 14.6  & 0.24 & 0.20 & 0.36 & 0.26 \\
BGDN-192   & 15.2  & 0.26 & 0.22 & 0.37 & 0.25 \\
VA-Fact  & 17.4  & 0.41 & 0.24 & 0.62 & 0.29 \\
VA-Hype  & 29.5 & 0.58 & 0.44 & 0.70 & 0.36 \\
VA-Mean      & 96.5  & 1.01 & 0.77 & 1.01 & 0.54 \\
HiFiC      & 693.1 & 2.10 & 1.75 & 4.53 & 2.65 \\
\hline
\end{tabular}
\end{table}

JPEG2000 lower mean values seen in Fig. \ref{fig:overall_psnr_ssim} are translated in substantial lower execution times when compared to the learning models. We can see that the better quality results obtained with the VA and HiFiC models come with a higher cost in complexity. The results also show that decompression is usually more time-consuming than compression. Considering the learning-based models, on average, decompression takes 49.1\% more time on a CPU and 16.12\% on a GPU, when compared to compression. This brings attention to resource-constrained devices, which usually run decompression operations much more often than compression. Therefore, solutions that reduce the complexity of decompression are highly necessary. We can also see that there is a direct relationship between model size and processing time, which is expected because larger models lead to more operations. The Pearson correlation between these two metrics ranged between 0.95 and 0.99. We can also conclude that running the learning-based models on a GPU led to modest time savings. On average, savings of 19.2\% (1.26$\times$ speedup) and 37\% (1.64$\times$ speedup) were observed on the compression and decompression times respectively. This shows that GPUs might not be an advisable solution for efficient compression on resource-constrained devices, since these units greatly increase chip cost and power dissipation.

\begin{figure}[!t]
    \centering
    \includegraphics[width=0.95\linewidth]{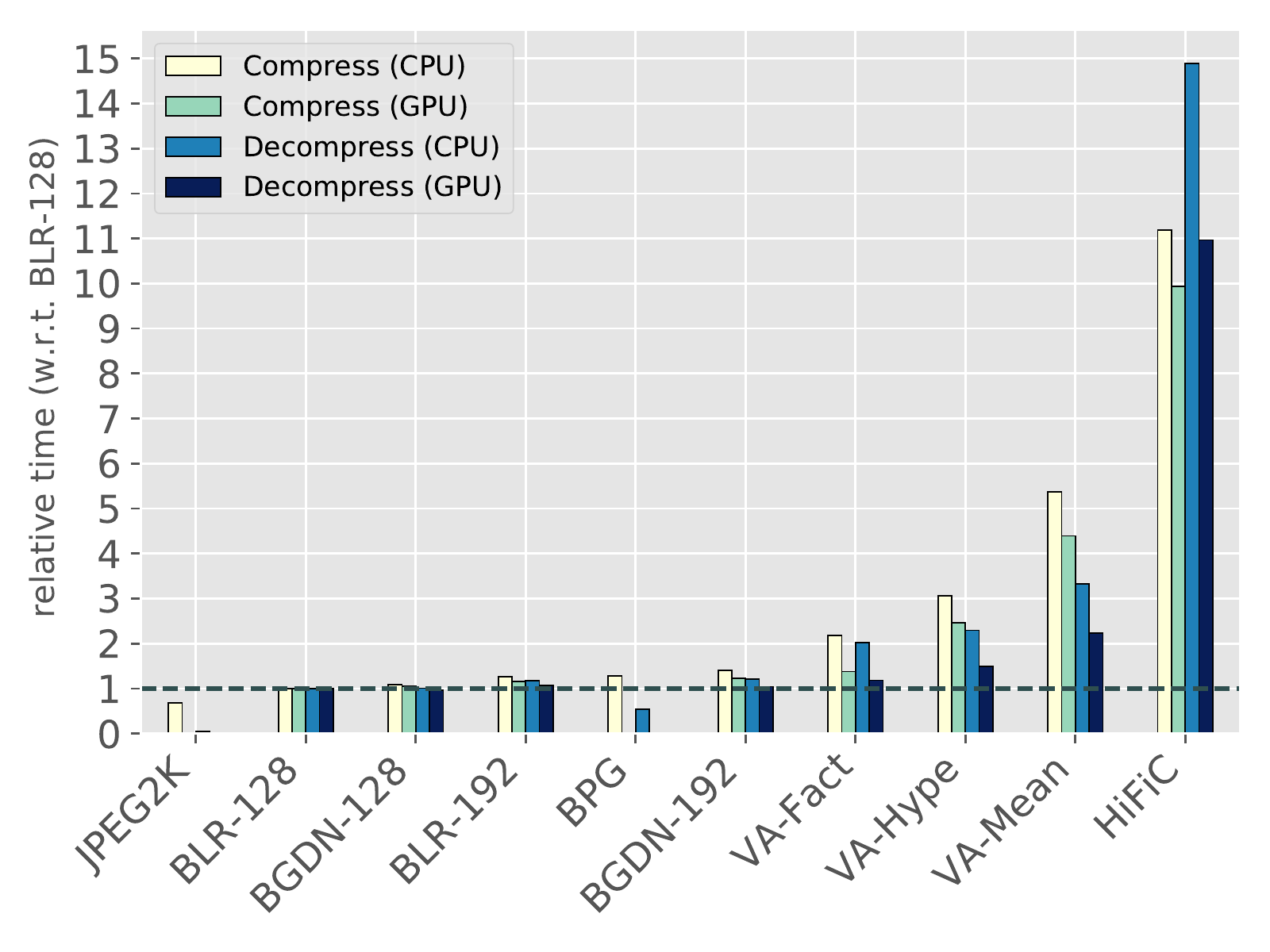}
    \caption{Processing time ratios using BLR-128 as baseline for comparison running on CPU and GPU platforms.}
    \label{fig:relative_times}
        \vspace{-0.5cm}
\end{figure}

Fig. \ref{fig:relative_times} shows the relative processing time, using the least complex BLR-128 model as baseline for comparison. The charts show that complexity scales significantly in the HiFiC models, taking 11.05 and 15.1$\times$ more time on CPUs to compress and decompress images respectively. On GPUs, these values reach factors of 9.72$\times$ for compression and 11.04$\times$ for decompression. The VA family presented a moderate increase, taking up to 5.32$\times$ more time to compress images on CPUs with the VA-Mean model (which has also the largest network of the three). The BPG codec presents lower execution times, but it is more time-consuming for compression on CPUs compared to JPEG2000 and BLR-128. Lastly, we can observe that the BGDN family has a very low computing increment ($<$1.37$\times$) while also leading to better quality results. A final conclusion of this analysis is that the lower execution times for JPEG2000 is seen by its comparison with the baseline model, leading to 1.46$\times$ and 30$\times$ reduced time in compression and decompression, respectively.

\section{Conclusions}
\label{sec:conclusions}

This paper presented a detailed analysis of a recent paradigm of learning-based image compression, i.e. using CNNs to compress images, as opposed to the traditional methods. We analyzed 8 different models from different works and compared them to JPEG2000  and BPG in terms of different visual quality metrics -- PSNR, MS-SSIM and LPIPS. The models were also compared in terms of execution time while running in a CPU and a GPU. We found that learning-based solutions have a great potential for exploration, since this new approach yields a different range of architectures. On the downside, a great complexity penalty is still present, requiring solutions to decrease complexity and to accelerate these CNN models.

\vfill\pagebreak
\bibliographystyle{IEEEbib}
\bibliography{ref}


\end{document}